\documentclass{article}
\usepackage{graphicx} % Required for inserting images
\usepackage{amsmath}
\usepackage{booktabs}
\usepackage{geometry}
\usepackage{tikz} % Required for inserting images
\usepackage{url}
\usepackage{hyperref}
\usepackage{listings}
\usepackage[]{mdframed} % text inside of a box
\usepackage{authblk}
\usepackage{float}

\title{Helsinki Speech Challenge 2024}
\author[1]{Martin Ludvigsen}
\author[2]{Elli Karvonen}
\author[2]{Markus Juvonen}
\author[2]{Samuli Siltanen}

\affil[1]{Department of Mathematical Sciences, Norwegian University of Science and Technology.}
\affil[2]{Department of Mathematics and Statistics, University of Helsinki.}

\affil[*]{Email: hsc2024@helsinki.fi}

\date{\today}

\begin{document}

\maketitle

\tableofcontents

\section{Introduction}
We present the Helsinki Speech Challenge 2024 (HSC2024). We invite researchers and scientists to try their speech enhancement and audio deconvolution algorithms on our real world data. Participants will be provided with our recorded dataset, consisting of paired clean and corrupted speech audio samples affected by our recording setup. The challenge is to develop innovative methods to effectively recover the clean audio from these corrupted recordings.

The main goal of the challenge is threefold. Firstly, we aim to challenge the prevailing practice of synthesizing paired training data, which is common in audio processing. To our knowledge, large-scale datasets featuring real-world, convolved audio data in challenging conditions, such as those presented in this dataset, are uncommon.

Secondly, we seek to bridge the gap between two seemingly disconnected research domains: inverse problems, which traditionally employ tools from applied and abstract mathematics, and speech enhancement, which has increasingly relied on machine learning in recent years \cite{mueller2012linear,speechEnhancement}.

Lastly, we aim to demonstrate the utility of speech enhancement for downstream tasks such as speech recognition. Conversely, we propose using speech recognition models as a means to quantify the performance of speech enhancement algorithms.

Following the tradition of the Helsinki data challenges, our data is divided into different difficulty levels. The group that is able to beat the most levels wins the challenge, and will be awarded a prize, as well as be invited to present their results at this years Inverse Days the 10. - 13. December in Oulu, Finland. 
%TODO: Do we want to add the prize? 

Even after the challenge is over, we still hope that this data will be used for research in the fields of speech enhancement and inverse problems.

\section{Speech Enhancement and Deconvolution}

Speech enhancement encompasses processes like denoising and defiltering to improve speech quality and intelligibility by reducing noise and filtering from recorded speech signals. It involves the extraction of clean speech from noisy and filtered mixtures, which are typically modeled as the sum of the desired speech, noise, and often a convolution with room impulse responses. This task is crucial in fields such as telecommunications, hearing aids, and automated speech recognition systems, where background conditions can significantly impair the signal. 

\subsection{Speech Enhancement as an Inverse Problem}
Mathematically, the general speech enhancement problem is to recover the speech signal of interest $x(t)$ given a recorded signal $y(t)$, following the mathematical model
\begin{equation}
    y(t) = A(x(t)) + u(t) + w(t),
\end{equation}
where $A$ is some possibly non-linear filter, $u(t)$ is some non-stationary signal that is not of interest and $w(t)$ is stationary noise. 

Recording systems can usually be assumed to be so-called Linear Time Invariant (LTI), in which case speech enhancement is the problem of deconvolution
\begin{equation}
    y(t) = (k * x)(t) + w(t),
\end{equation}
where $*$ denotes convolution, and $k$ is the so-called impulse response of the system.

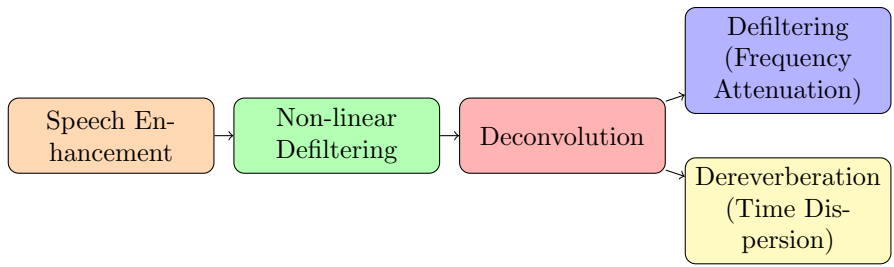
\begin{figure}[!htb]
\centering
\begin{tikzpicture}
% Define styles
\tikzset{box/.style={draw, minimum width=2.5cm, minimum height=1cm, text centered, text width = 2.5cm, rounded corners}}

% Nodes
\node[box, fill=orange!30] (enhancement) at (0,0) {Speech Enhancement};
\node[box, fill=green!30] (high) at (3,0) {Non-linear Defiltering};
\node[box, fill=red!30] (convolution) at (6,0) {Deconvolution};
\node[box, fill=blue!30] (filtering) at (9,1) {Defiltering (Frequency Attenuation)};
\node[box, fill=yellow!30] (reverb) at (9,-1) {Dereverberation (Time Dispersion)};

% Lines
\draw[->] (enhancement) -- (high);
\draw[->] (high) -- (convolution);
\draw[->] (convolution) -- (filtering);
\draw[->] (convolution) -- (reverb);

\end{tikzpicture}
\caption{Diagram showing some of the different levels of speech enhancement.}
\label{fig:speech}
\end{figure}

In speech processing and enhancement, the effect of impulse responses can be broadly categorized into two main types: frequency attenuation and temporal dispersion. A categorization of these is shown in Figure \ref{fig:speech}.

Frequency attenuation, generally termed filtering, refers to the consistent effect on the audio signal as it passes through a medium or recording equipment. This process modifies the signal's frequency content, often selectively amplifying or attenuating specific frequency ranges based on the properties of the medium. This combined effect of the medium and the recording equipment results in a filtered version of the original audio signal.

On the other hand, temporal dispersion responses, more commonly referred to as reverberation, involve the dispersal of the audio signal over time. This type of response results from the audio signal bouncing off various surfaces such as walls and objects within an environment, which leads to multiple delayed echoes. These echoes superimpose and decay gradually, creating a sense of space and depth in the audio. Unlike filtering, which primarily affects the frequency domain, reverberation impacts the temporal structure of the signal.

Both of these impulse responses can be modeled through convolution, represented as pointwise multiplication in the Fourier domain using the Fourier Transform (FT) or Short Time Fourier Transform(STFT). In the Fourier domain, this often leads to certain frequency coefficients being significantly reduced or eliminated due to the impulse response characteristics, causing a loss of information. This reduction makes it mathematically ill-posed, as reconstructing the original signal becomes infeasible without additional constraints or prior knowledge.

Informally, filters alter the spectrogram of a signal only in frequency, and reverb alters the spectrogram in time. Examples of this are shown in the experimentally obtained data in Figure \ref{fig:data1} and \ref{fig:data2}.

In real-world audio settings, the linear time-invariant (LTI) model can prove insufficient, as it fails to capture the various non-linear effects that can occur, as well as possible time-variances. These non-linear effects can include harmonic distortion, where higher frequency harmonics are generated by devices or media that do not respond linearly to the signal; saturation, where the signal amplitude is limited, leading to a flattening of audio peaks; and intermodulation distortion, where new frequencies are produced from the mixing of two or more different frequencies within the audio signal. Such complexities necessitate more sophisticated models to accurately describe and process real-world audio.

Compared to imaging, another common application of inverse problems, there are relatively few effective model-based priors for speech signals due to their inherent complexity. This scarcity of effective model-based priors has led to a greater reliance on data-driven approaches, which leverage large datasets to learn the underlying patterns and structures of speech signals, thereby providing more robust and adaptable solutions. Data-driven approaches also often avoid the modeling of non-linear effects, which can be quite challenging to model in practice. As a consequence of this, large quantities of speech and realistic noise data has been curated by the speech enhancement community. A common approach for learning is then to synthetically create paired data by convolving speech with known impulse responses, and mixing it with noise. While this is an effective, there is a possibility that it does not fully capture the complexities of realistic audio recording and both analog and digital signal processing. 

Additionally, obtaining reliable metrics for audio quality is notably challenging, due to the subjective nature of audio perception, which complicates the objective assessment of audio fidelity.

\subsection{The Mozilla DeepSpeech Model as a Quantitative Metric}

Inspired by the 2021 Helsinki Deblur Challenge \cite{HDC2021}, we propose using a speech recognition model as a metric to determine the quality of recovered audio. This approach is motivated by two main arguments.

Firstly, speech recognition serves as an effective quantitative metric because these models are designed to transcribe high-quality, clean speech audio into text. This characteristic ensures that if the audio quality is high, the model will perform well, while significant corruption in the audio will degrade the model’s performance, as the model has limited capacity. This is a point point we verify experimentally in Figure \ref{fig:CER}.

Secondly, speech recognition is a practical downstream task following speech enhancement. Enhanced speech is often transcribed in real-world applications, making it logical to evaluate speech enhancement techniques based on their effectiveness in improving speech recognition outcomes. Specifically, we propose that one approach to speech recognition on corrupted data is to first enhance the speech, then perform recognition, as opposed to training the speech recognition model directly on corrupted data.

Compared to traditional speech metrics such as PESQ (Perceptual Evaluation of Speech Quality) and STOI (Short-Time Objective Intelligibility) \cite{PESQ,STOI}, which can be applied to audio signals, our proposed setup requires a method to compare text outputs. We propose using the Character Error Rate (CER) for this purpose.

CER is calculated by comparing the characters in the true and recovered text, and is defined as:
\begin{equation}
    \text{CER} = \frac{S + D + I}{N},
\end{equation}
where $S$, $D$ and $I$ correspond to the amount of substitutions, deletions and insertions, respectively, between the reference and proposed text, and $N$ is the total number of characters.

To ensure a fair comparison, we propose several pre-processing steps before calculating CER. First, we remove spaces and other formatting, as well as map everything to lower case to prevent artificially high CER values from minor formatting differences, such as "Bestfriend" versus "best friend." Additionally, we substitute common variations between British and American English to avoid discrepancies caused by the use of the American English DeepSpeech model on British English text. For example, the string "colour" is replaced with "color". 

We experimentally found that using CER is more robust than metrics such as Word Error Rate (WER), which can be sensitive to small changes, like the string examples previously mentioned. 

In cases where the transcribed text is empty, we set the Character Error Rate (CER) to $1$, indicating that the transcription is completely incorrect. Consequently, the CER is slightly biased away from $1$, as nearly any non-empty string that contains some characters from the reference string will yield a CER lower than $1$.

We chose to use the Mozilla DeepSpeech \cite{DeepSpeech_article} model for speech recognition due to its high performance and convenient Python interface. In our evaluations, the DeepSpeech model performed exceptionally well on our clean training data, achieving a median CER of $0$.

%One drawback of the DeepSpeech model is its computational cost. On a standard laptop, the real-time factor (RTF) was sometimes larger than 1, meaning that transcribing text took longer than the duration of the audio file itself. This could be problematic for large-scale computations and parameter tuning scenarios.

\section{Clean Data}\label{sec:clean_data}
Although there exists many datasets that are commonly used for speech enhancement tasks, like the LibriSpeech dataset \cite{LibriSpeech} and the WSJ0-2mix \cite{WSJ0-2mix} dataset, we wanted to create new data for the data challenge. Given the challenge of obtaining real speech data with annotated text, we instead obtain speech data through
OpenAI's text-to-speech model tts-1 \cite{OpenAI_tts}, which generates speech data from text. Modern text-to-speech models like tts-1 create high quality audio that is essentially indistinguishable from realistic audio, while also creating audio that is virtually noise-free unlike the mentioned datasets which have some low quality data. 

The tts-1 model contains six built-in voices, that were "Alloy", "Echo", "Fable", "Onyx", "Nova", and "Shimmer". The model tts-1 produces audios of 24kHz which were down-sampled to 16kHz, and had their audio levels normalized.

The text samples used to generate the audio samples were from the following books from Project Gutenberg:
\begin{itemize}
    \item The Arrow of Gold: A Story Between Two Notes, J. Conrad \cite{Conrad_gutenberg}
    \item Emma, J. Austen \cite{Austen_gutenberg}
    \item Harriet and the Piper, K. Norris \cite{Norris_gutenberg}
    \item The Turn of the Screw, H. James \cite{James_gutenberg}
    \item The Jungle Book, R. Kipling \cite{Kipling_gutenberg}
    \item The King in Yellow, R. W. Chambers \cite{Chambers_gutenberg}
    \item The Adventures of Pinocchio, C. Collodi \cite{Collodi_gutenberg}
    \item A Room with a View, E. M. Forster \cite{Forster_gutenberg}
    \item The Rosary, F. L. Barclay \cite{Barclay_gutenberg}
    \item The Time Machine, H. G. Wells. \cite{Wells_gutenberg}
\end{itemize}
The samples were sampled sentences from the above books. Each sentence had more than five words. We attempted to avoid sentences that contained names of people, honorifics such as Mr., Ms. Mrs., and Dr., and symbols such as quotation marks, apostrophes, and brackets. 

The text samples were shuffled, and then divided by a number of words in a sentence. Short sentences had 6-10 words, medium length sentences had 11-20 words and long sentences had more than 20 words. Furthermore we divided samples to 10 sets having each 650 samples. In the each set 319 samples were short, 306 samples were medium length and the rest 25 samples were long. The voice distribution of each set were following one. Both Alloy and Echo had 109 samples, and the rest had 108 samples each. 

Furthermore, we added some padding around the signals, as we noticed that the Deepspeech model was somewhat sensitive to the lack of padding around signals. For all data, we added half a second at the start and end, and for all data that would later be used for convolution experiments, we added $5$ seconds at the end.

To remove any clear outliers for initial sets, we further passed the data through the DeepSpeech model, and calculated the CER between the transcribed and true text. We then removed $5\%$ of the data that had the worst score. Most of the poor results were caused by proper names of people and locations, as well as text that was in different languages and dialects. This was also done to ensure that the DeepSpeech model was a suitable metric on this data. After this process, the Mean CER between transcribed and true text was between $0.005$ and $0.009$ for the different sets. 

The 10 preprocessed sets form the base of the final levels that we used. 
The final distributions of sentence lengths and voices as well as number of used samples in each level is presented in Appendix Table \ref{tab:data_info_clean}.

% TODO: Explain this further, that is, the point was we wanted to remove outliers

\section{Experiment Setup and Data}

To capture real-world filtered and reverberated audio, we propose two distinct experimental setups.

\begin{figure}[!htb]
    \centering
    \includegraphics[width=0.99\textwidth]{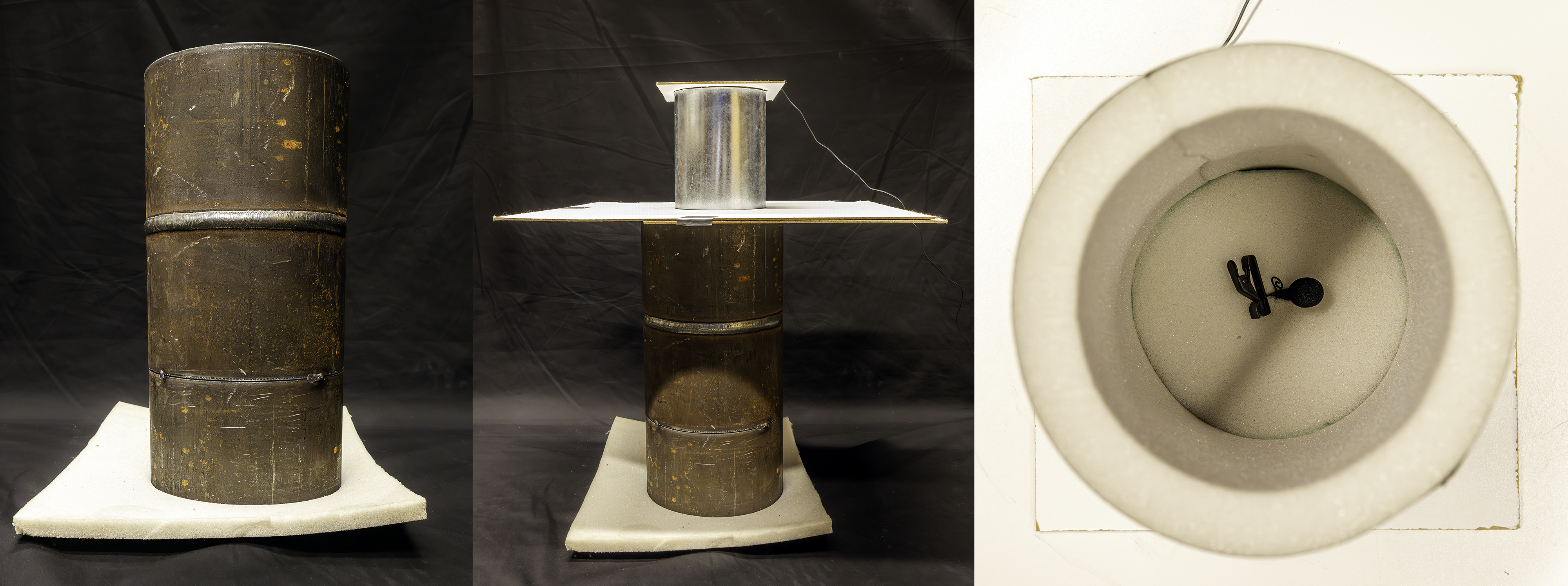}
    \caption{The first setup. Left: Side-view of the lower metal pipe. Middle: Side-view of the pipe with the second piece of pipe on top. Right: View from the bottom inside the second part of the pipe that has the microphone attached inside.  }
    \label{fig:lab1}
\end{figure}

\begin{figure}[!htb]
    \centering
    \includegraphics[width=0.99\textwidth]{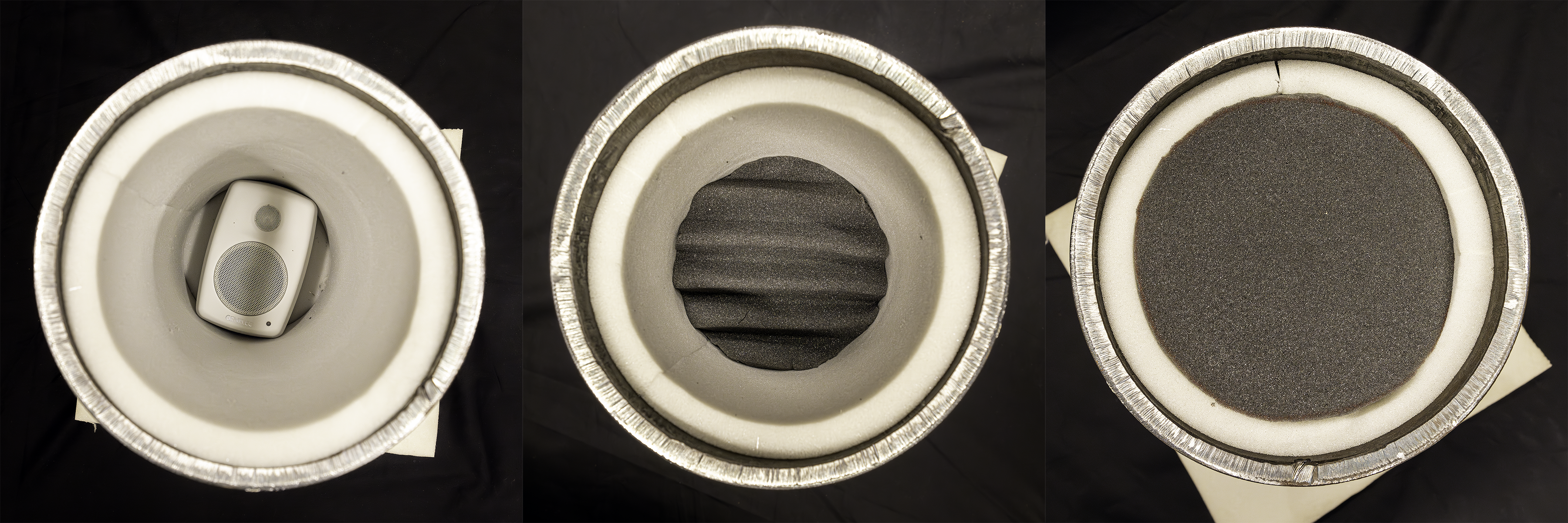}
    \caption{The first setup. Left: View from top into the lower pipe that has the speaker on the bottom. Middle: View from the top with some material in the lower tube. Right: View from the top with more material in the lower tube. }
    \label{fig:lab2}
\end{figure}

\begin{figure}[!htb]
    \centering
    \includegraphics[width=0.99\textwidth]{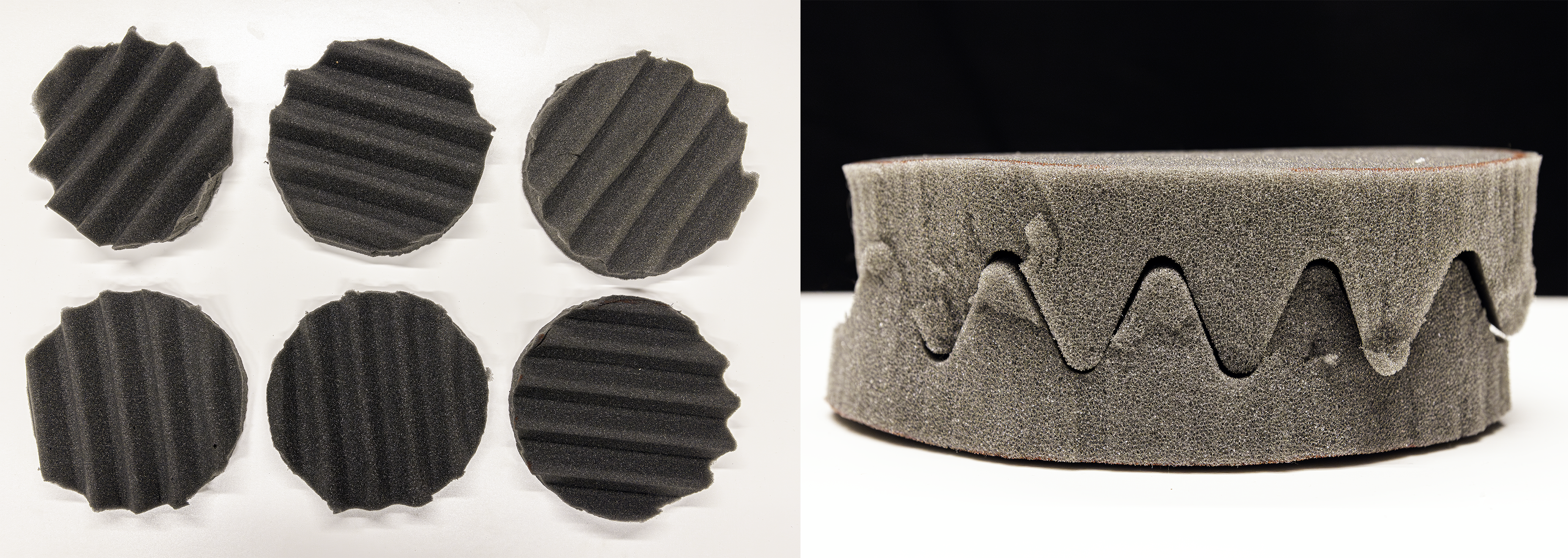}
    \caption{Acoustic foam that was used as material between the speaker and microphone. The foam had a density of $30$ kg/m$^3$. }
    \label{fig:foam}
\end{figure}

\begin{figure}[H]
    \centering
    \includegraphics[width=0.99\textwidth]{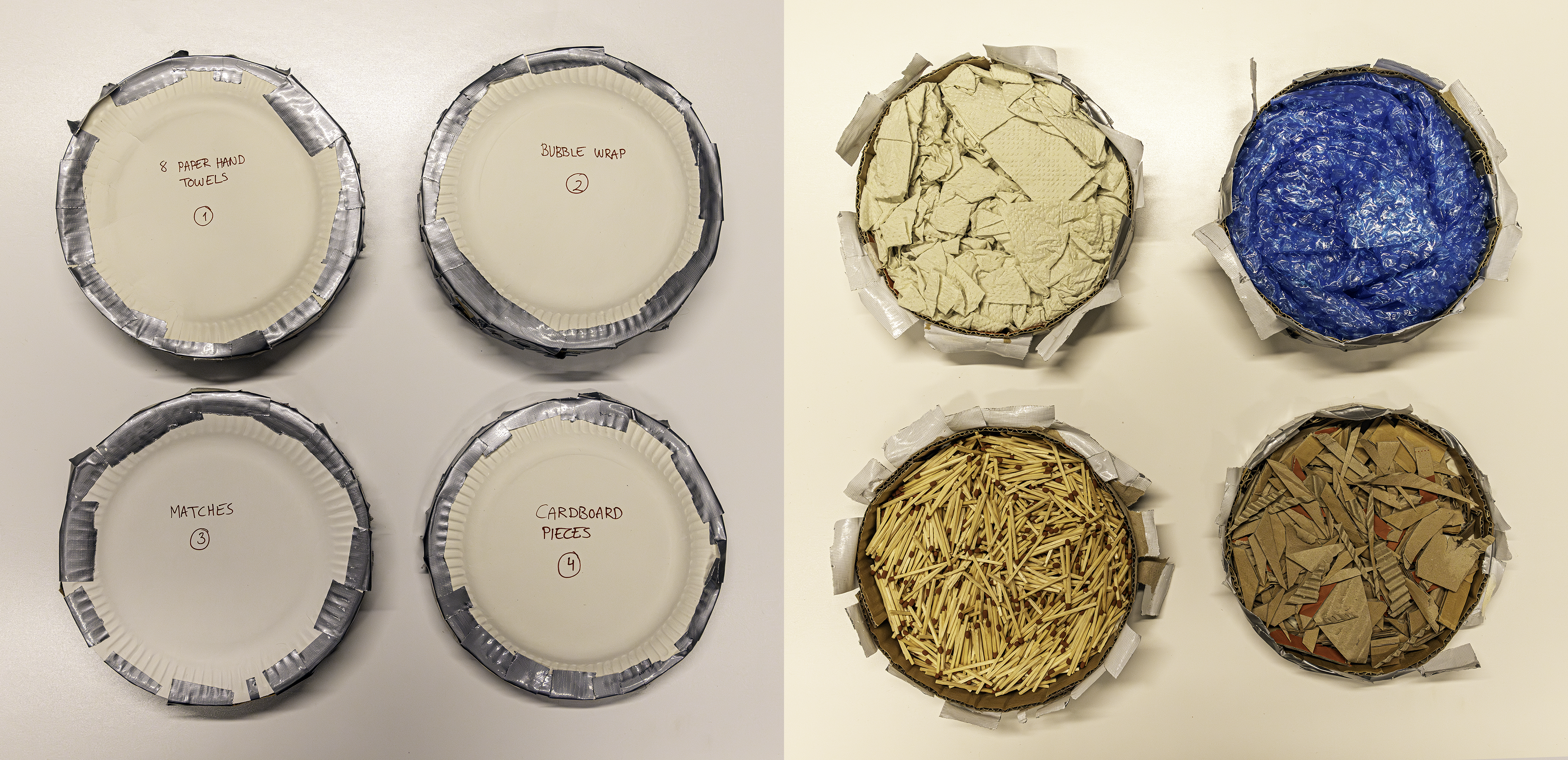}
    \caption{Discs made out of two paper plates filled with different materials (paper towels, bubble wrap, matches, cardboard pieces) and taped together were also used between the speaker and microphone. }
    \label{fig:discs}
\end{figure}

The first setup, termed the "filter experiment," involves placing a speaker and a microphone at opposite ends of soundproof tubes padded with a 20 mm layer of foam on the inside. The speaker is placed inside the base of the larger bottom tube, which is 519 mm in height and has a 228 mm inside diameter. The microphone is placed inside the top of the smaller tube which has a height of 200 mm and a 160 mm diameter. The larger tube is then filled with progressively more additional soundproof foam and other materials such as paper towels, bubble wrap, matches and cardboard pieces, to increase the ill-posedness of the filter. Two medium-density fibreboards with suitably sized holes were placed between the two tubes of varying diameter to seamlessly join them together and stabilize the setup. When doing recordings, we also placed a blanket on top of the entire setup, in an attempt to muffle outside noise. The audio recorded at the microphone end is thus a filtered version of the original signal. Additionally, the speaker and microphone inherently filter the signal and introduce recording noise. Since achieving complete soundproofing was not feasible, some ambient noise will also inevitably seep into the recording. The first setup is photographed in Figure \ref{fig:lab1} and \ref{fig:lab2}, and the materials are captured in Figure \ref{fig:foam} and \ref{fig:discs}.

\begin{figure}[!htb]
    \centering
    \includegraphics[width=0.99\textwidth]{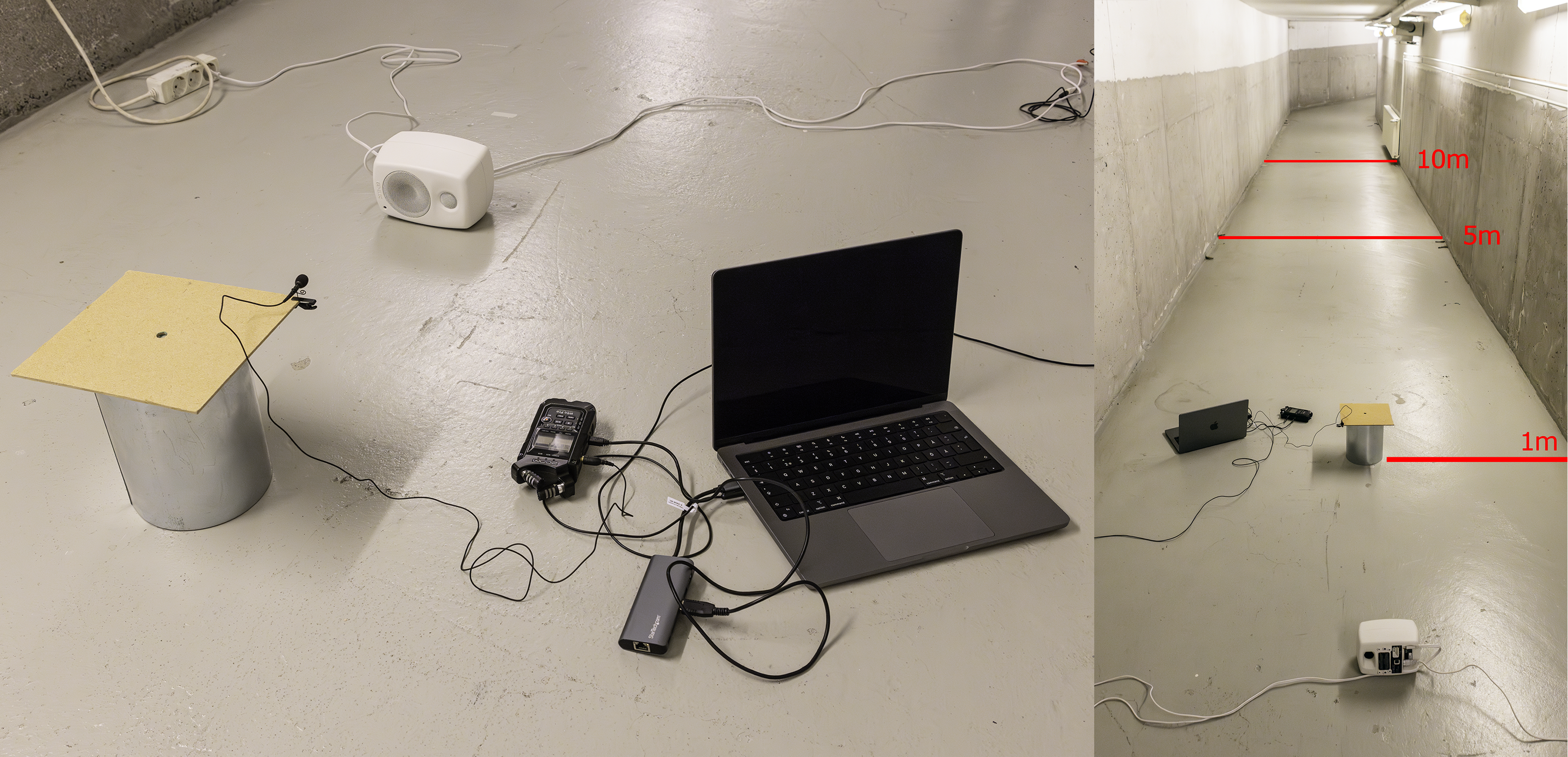}
    \caption{The second setup a.k.a. the "reverb experiment". }
    \label{fig:tunnel}
\end{figure}

\begin{figure}
\centering
\begin{tikzpicture}[scale=1.2, every node/.style={transform shape}]
    % Soundproof Pipe
    \draw[thick] (0,0) rectangle (10,4);

    % Medium
    \draw[fill=blue!20] (4,0) rectangle (6,4) node[pos=.5] {Medium};

    % Speaker (audio source)
    \draw[fill=red!30] (3,2) circle [radius=0.5];
    \node at (3,0.5) {Speaker}; % Label below the speaker

    % Microphone
    \draw[fill=green!30] (7,2) -- (7.5,2.5) -- (7.5,1.5) -- cycle;
    \node at (7.25,1) {Microphone}; % Label below the microphone

    % Ambient Noise Source
    \draw[fill=red!30] (9,2) circle [radius=0.5];
    \node at (9,1) {Noise}; % Label below the ambient noise source

    % Computer
    \node (computer) at (2, 5) [draw, rectangle, minimum width=2.5cm, minimum height=1.5cm, align=center] {Computer};

    % Field Recorder
    \node (recorder) at (8, 5) [draw, rectangle, minimum width=2.5cm, minimum height=1.5cm, align=center] {Field recorder};

    % Connections
    \draw[<->] (computer) -- node[midway, above] {USB} (recorder);
    \draw[<-] (recorder) -- node[midway,right] {Audio cable} ++(0, -3)  --  (7.25, 2.0); % Connection to microphone
    \draw[->] (computer) -- ++(0, -3) node[midway,left] {Audio cable} --  (3, 2); % Connection to speaker

    % Labels
    \node at (5, -0.5) [align=center] {Room};

\end{tikzpicture}
\caption{Illustration of experimental setup. In the "filter experiment", the room is a roughly soundproof tube, and the medium is the material in the tube. In the "reverb" experiment, the room is a hallway, and the medium is simply the distance between the speaker and microphone. }
\label{fig:setup}
\end{figure}
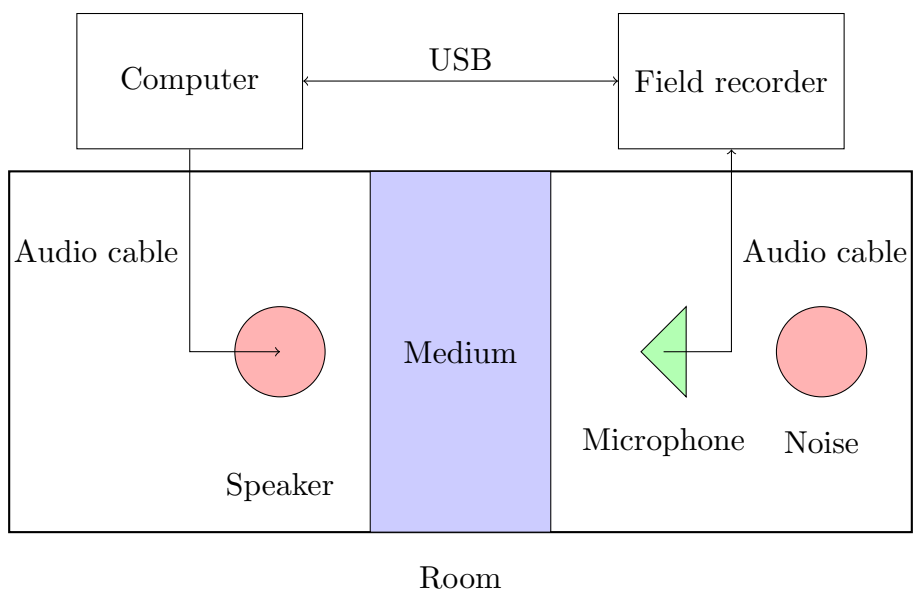

The second setup, referred to as the "reverb experiment," is similar to the filter experiment but takes place in a long, enclosed, underground hallway instead of a tube. In this environment, the audio signal bounces off the walls, resulting in heavily reverberated recordings. To progressively increase the level of reverberation, the microphone is placed at varying distances from the speaker, see Figure \ref{fig:tunnel}. The hallway also had a significant amount of roughly stationary ambient noise coming from ventilation systems and other machinery. Figure \ref{fig:setup} illustrates the experimental setups.


\newpage
\subsection{Recording Equipment}

A goal of the challenge was to use relatively low-end equipment to simulate realistic recording settings and introduce actual recording noise. Specifically, we chose to use a field recorder instead of an audio interface, which typically has higher recording noise, to better emulate practical, real-world recording conditions. 

The equipment used:
\begin{itemize}
    \item Genelec 8010 AW speaker.\cite{Speaker_Genelec} 
    \begin{itemize}
        \item \textbf{Frequency Response}: 67 Hz – 25 kHz (-6 dB)
        \item \textbf{Maximum SPL (Sound Pressure Level)}: 96 dB
        \item \textbf{Amplifier Power}: 25 W for Bass (Class D) and 25 W for Treble (Class D)
        \item \textbf{Dimensions}: 195 x 121 x 115 mm (with Iso-Pod)
        \item \textbf{Weight}: 1.5 kg
        \item \textbf{Connections}: Balanced XLR input.
    \end{itemize}
    \item Røde RØDELink LAV lavalier microphone.\cite{microphone} 
    \begin{itemize}
        \item \textbf{Acoustic Principle}: Permanently Polarized Condenser
        \item \textbf{Polar Pattern}: Omnidirectional
        \item \textbf{Frequency Range}: 20 Hz – 20 kHz
        \item \textbf{Output Impedance}: 3k$\Omega$
        \item \textbf{Signal to Noise Ratio}: 67 dB
        \item \textbf{Equivalent Noise Level (A-weighted)}: 27 dB
        \item \textbf{Maximum SPL}: 110 dB SPL (1 kHz @ 1\% THD)
        \item \textbf{Sensitivity}: -33.5 dB
        \end{itemize}
    \item A Zoom H4N Field recorder.\cite{recorder} \begin{itemize}
    \item \textbf{Input Impedance}: 2 k$\Omega$ or more
    \item \textbf{Input Gain (external microphone)}: - 16 dB to +51 dB
    \end{itemize}
\end{itemize}

As opposed to synthetically creating corrupted audio, a physical setup like this has some side effects. Firstly, the input and output audio files are not aligned, as there is some delay in the recording setup. In all audio files, there is a delay of up to half a second between the input and output files.

Secondly, the technical components of the setup are not perfect, and there are some recording artefacts in the recordings. Most notably, there is the occasional pop and crackle, as well as some segments where no audio was recorded. We have made no effort to remove or rerecord audio clips that had noticeable recording glitches.

Thirdly, while some effort was made to ensure that there were little to no noise from outside sources, there are some clips that are contaminated by outside noise. All data was recorded on the Kumpula campus of the University of Helsinki over several days in the afternoon, and we rerecorded data in the case where audio was clearly corrupted by outside noise. 

Lastly, the signal to noise ratio, as well as the presence of audio that is clipping, might vary for the different levels. Throughout the experiments, the gain level was set manually to be so that the peak record volume was around $-6$ dB on the recorder, though some levels have slightly less and more gain than this. 

\subsection{The Recorded Data}
\begin{figure}[!htb]
\centering
\begin{tikzpicture}
% Define styles
\tikzset{
    box/.style={draw, , minimum height=1cm, text centered},
    double color/.style={draw, minimum height=1cm, text centered, 
    fill=none, path picture={
      \fill[blue!30] (path picture bounding box.south west) -- (path picture bounding box.north west) -- (path picture bounding box.south east) -- cycle;
      \fill[yellow!30] (path picture bounding box.south east) -- (path picture bounding box.north west) -- (path picture bounding box.north east) -- cycle;
    }}
}
\node[box, fill=blue!30] (Box1) at (0,0) {T1 L1};
\node[box, fill=blue!30] (Box2) at (1.5,0) {T1 L2};
\node[box, fill=blue!30] (Box3) at (3,0) {T1 L3};
\node[box, fill=blue!30] (Box4) at (4.5,0) {T1 L4};
\node[box, fill=blue!30] (Box5) at (6,0) {T1 L5};
\node[box, fill=blue!30] (Box6) at (7.5,0) {T1 L6};
\node[box, fill=blue!30] (Box7) at (9,0) {T1 L7};

\node[box, fill=yellow!30] (Box8) at (0.0,-2) {T2 L1};
\node[box, fill=yellow!30] (Box9) at (1.5,-2) {T2 L2};
\node[box, fill=yellow!30] (Box10) at (3,-2) {T2 L3};
\node[double color] (Box11) at (12,0) {T3 L1};
\node[double color] (Box12) at (13.5,0) {T3 L2};
\end{tikzpicture}
\caption{The different levels of the data challenge. There are $7$ filtering levels,  $3$ reverb levels and $2$ combined levels. The specifics of the levels are explained in Table \ref{tab:exp}.}
\label{fig:ex_data}
\end{figure}
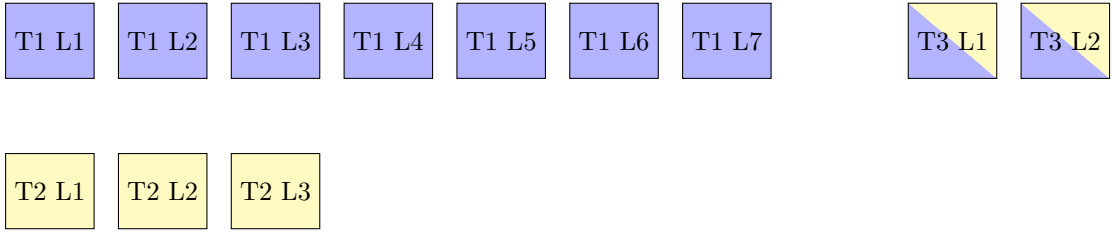

\begin{figure}[!htb]
    \centering
    \includegraphics[width=0.99\textwidth]{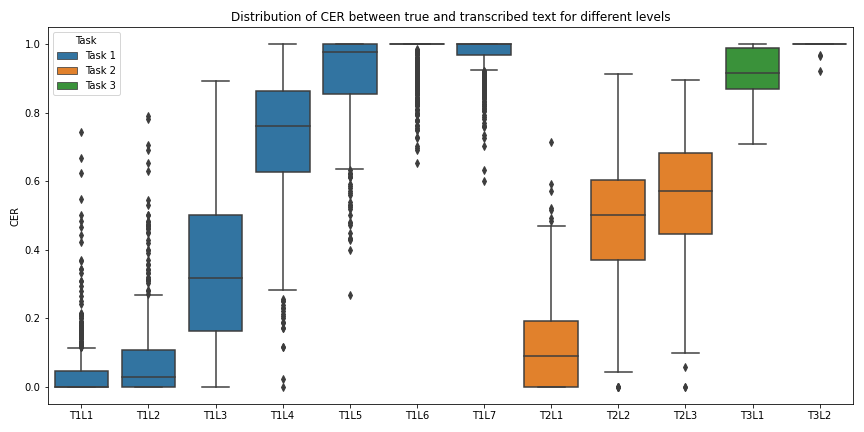}
    \caption{Distribution of CER values between true and transcribed recorded data for different levels. We note that the CER of the clean data are extremely low, indicating that the clean data consists of high-quality speech, and that CER  values tend to increases with the level of corruption. This showcases that while the DeepSpeech model is somewhat robust, it fails at transcribing corrupted audio. }
    \label{fig:CER}
\end{figure}

\begin{figure}[!htb]
\centering
\begin{tabular}{ccc}
\hline
\textbf{ID} & \textbf{Clean Data} & \textbf{Recorded Data} \\
\hline
T1L1 & \includegraphics[width=0.42\textwidth]{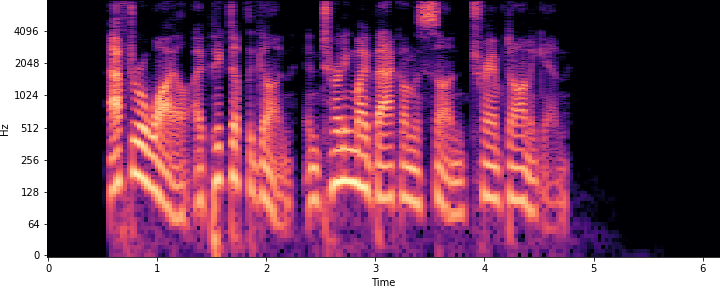} & \includegraphics[width=0.42\textwidth]{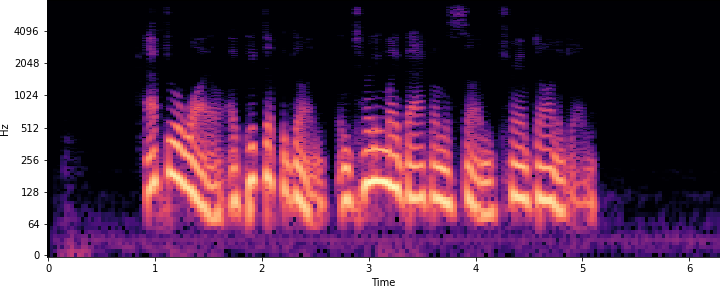} \\
\hline
T1L2 & \includegraphics[width=0.42\textwidth]{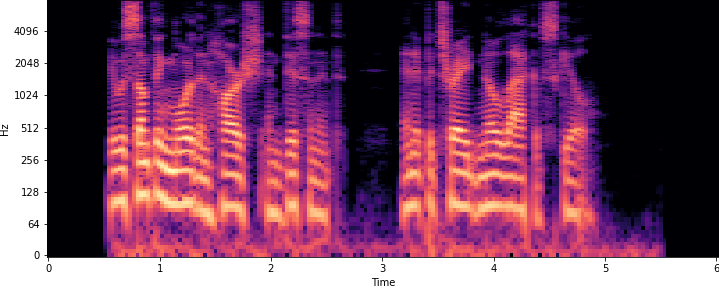} & \includegraphics[width=0.42\textwidth]{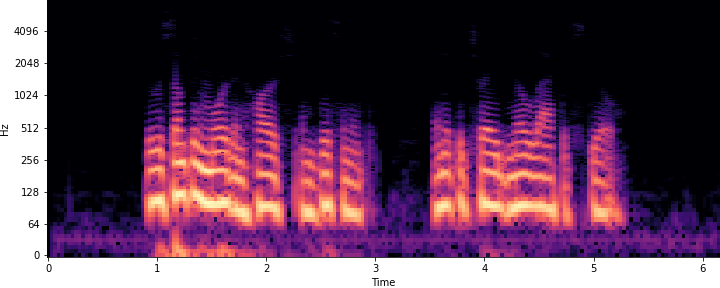} \\
\hline
T1L3 & \includegraphics[width=0.42\textwidth]{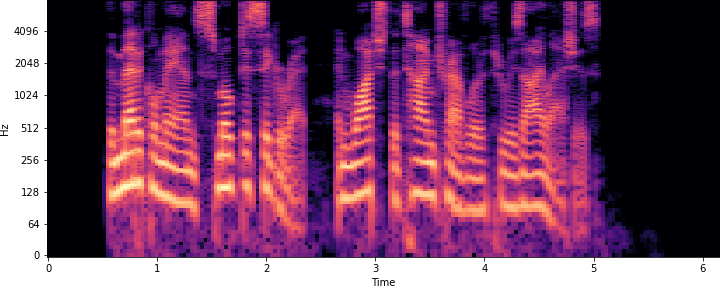} & \includegraphics[width=0.42\textwidth]{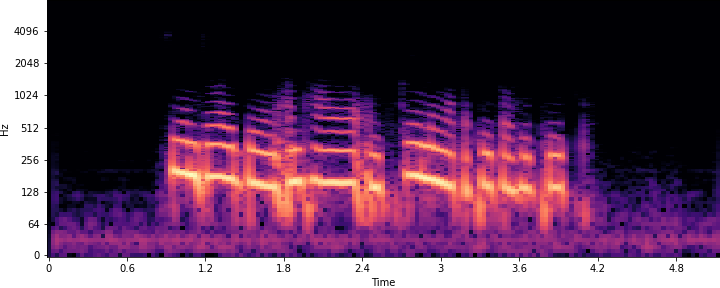} \\
\hline
T1L4 & \includegraphics[width=0.42\textwidth]{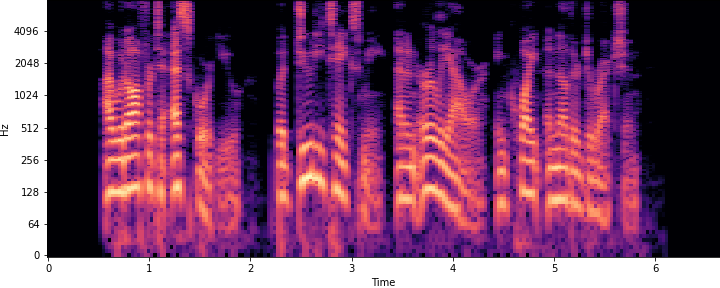} & \includegraphics[width=0.42\textwidth]{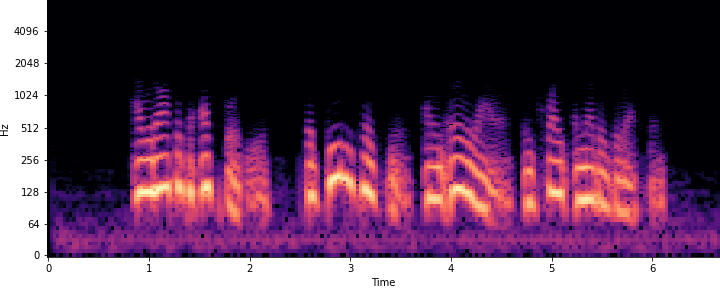} \\
\hline
T1L5 & \includegraphics[width=0.42\textwidth]{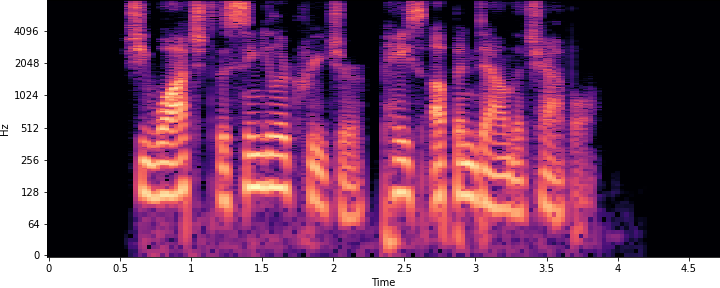} & \includegraphics[width=0.42\textwidth]{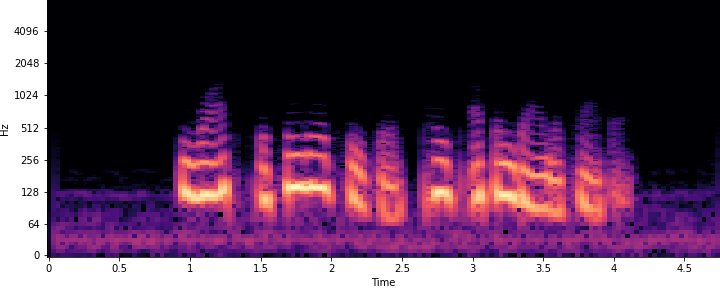} \\
\hline
T1L6 & \includegraphics[width=0.42\textwidth]{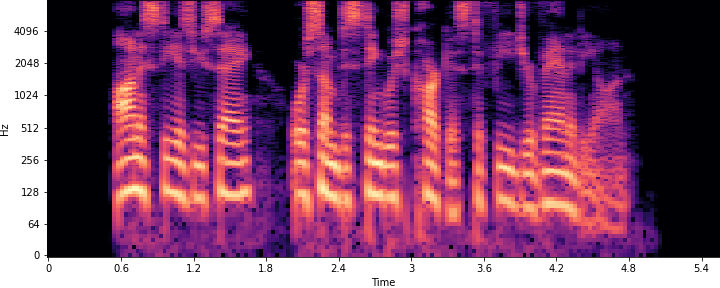} & \includegraphics[width=0.42\textwidth]{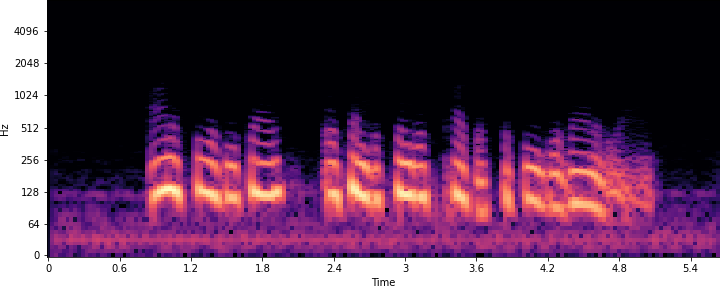} \\
\end{tabular}
\caption{Part 1 of Spectrogram comparison between Clean and Recorded Data for different levels.}
\label{fig:data1}
\end{figure}

\begin{figure}[!htb]
\centering
\begin{tabular}{ccc}
\hline
\textbf{ID} & \textbf{Clean Data} & \textbf{Recorded Data} \\
\hline
T1L7 & \includegraphics[width=0.42\textwidth]{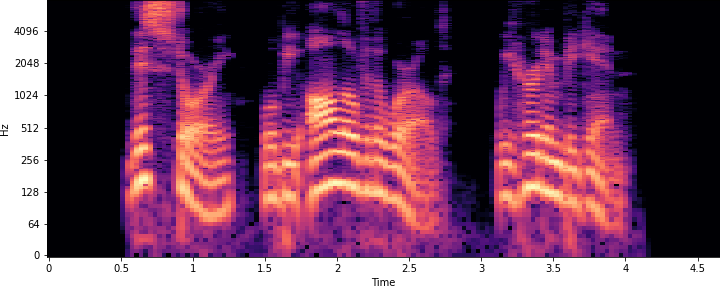} & \includegraphics[width=0.42\textwidth]{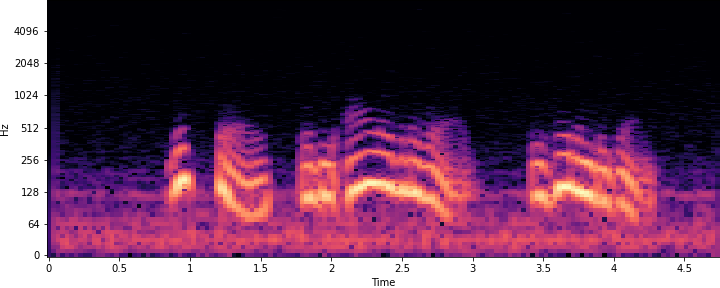} \\
\hline
T2L1 & \includegraphics[width=0.42\textwidth]{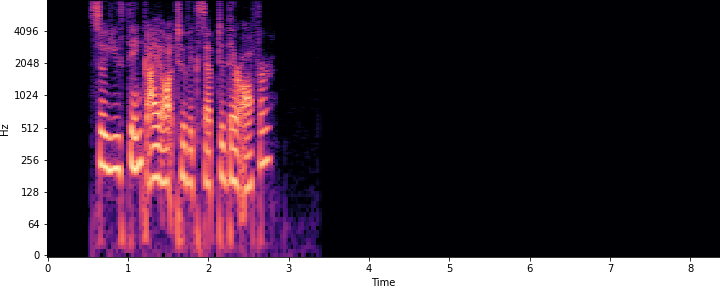} & \includegraphics[width=0.42\textwidth]{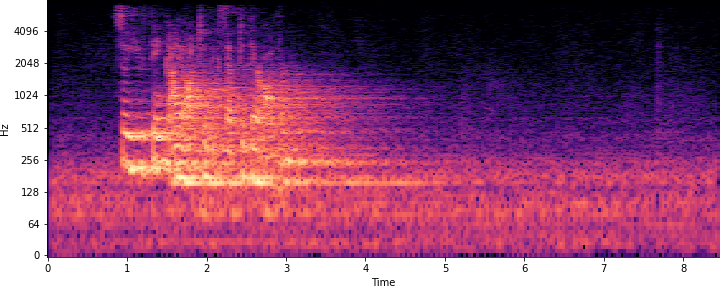} \\
\hline
T2L2 & \includegraphics[width=0.42\textwidth]{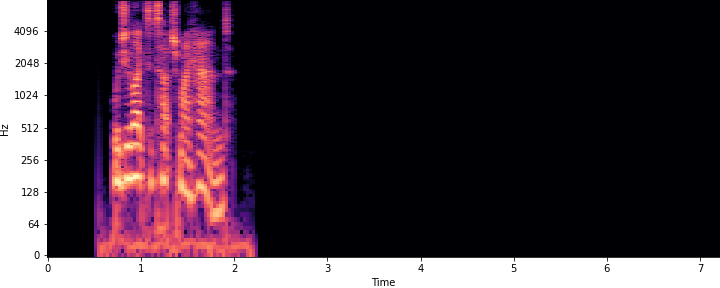} & \includegraphics[width=0.42\textwidth]{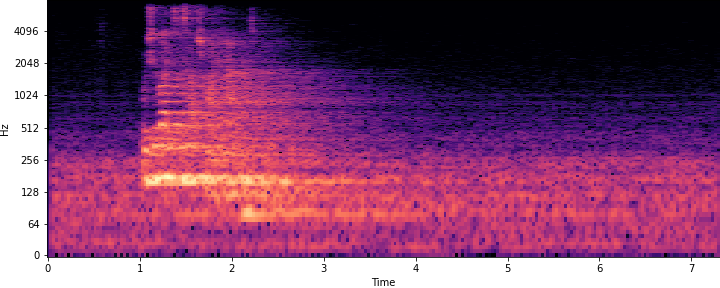} \\
\hline
T2L3 & \includegraphics[width=0.42\textwidth]{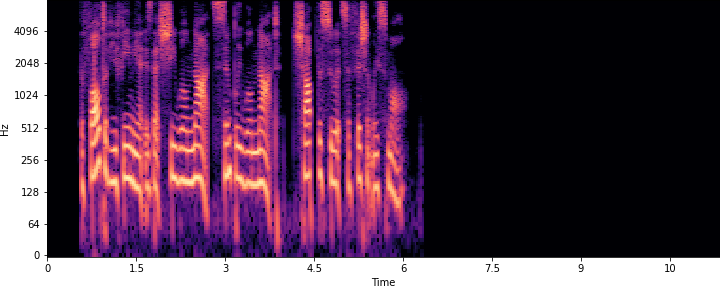} & \includegraphics[width=0.42\textwidth]{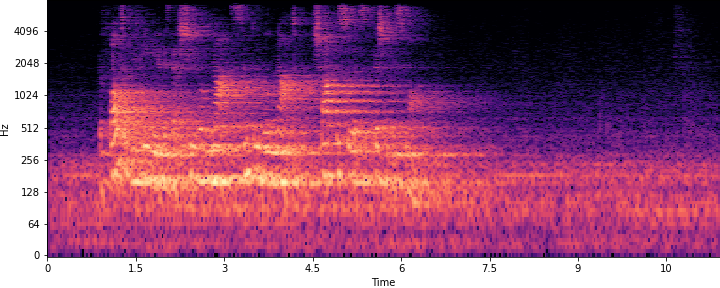} \\
\hline
T3L1 & \includegraphics[width=0.42\textwidth]{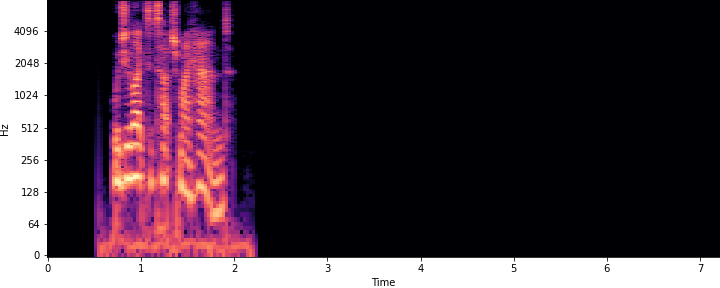} & \includegraphics[width=0.42\textwidth]{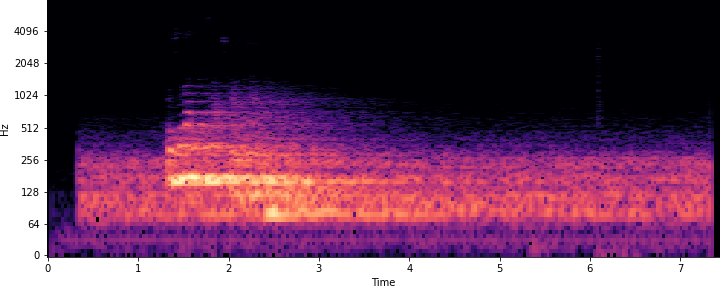} \\
\hline
T3L2 & \includegraphics[width=0.42\textwidth]{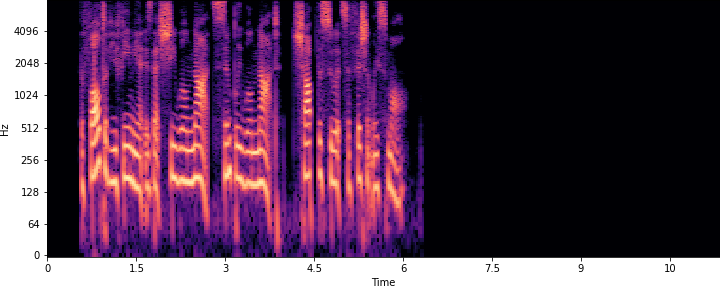} & \includegraphics[width=0.42\textwidth]{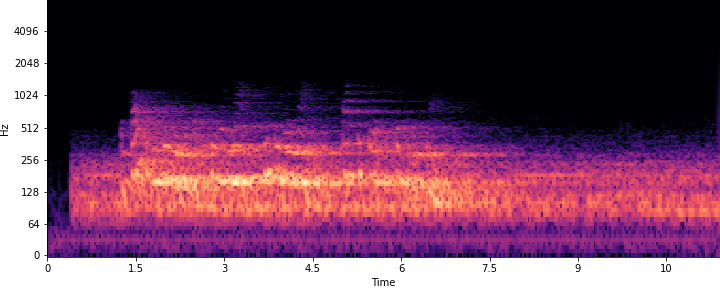} \\
\hline
\end{tabular}
\caption{Part 2 of Spectrogram comparison between Clean and Recorded Data for different levels.}
\label{fig:data2}
\end{figure}

The synthesized speech data was first divided into $10$ different sets as explained in Section \ref{sec:clean_data}. We then designed $7$ filtering experiments which we call task $1$, as well as $3$ reverb experiments, which we call task $2$. For the two last reverb experiments, we further rerecorded this data with the filter experiment setup, which we call task $3$. The different data levels is thus identified by a string on the form \texttt{TXLY}, where \texttt{X} refers to the task, and \texttt{Y} refers to the level. 
A figure illustrating the different levels is shown in Figure \ref{fig:ex_data}, the physical experiment parameters is shown in Table \ref{tab:exp}. 
We also ran the recorded data through the DeepSpeech model and calculated CER values. The result as well as some other statistics is shown in Table \ref{tab:data}, and in Figure \ref{fig:CER}. The spectrogram of example files for all levels are shown in Figure \ref{fig:data1} and \ref{fig:data2}. 

Because of a recording equipment failure, an additional $11$ files were removed from the \texttt{T1L1} data. Furthermore, because of some practical constraints for Task 2 and Task 3 we could only record roughly half the amount of files compared to Task 1, meaning that these tasks are additionally difficult because the amount of speech signal is less. 

\begin{table}[!htb]
\centering
\begin{tabular}{@{}lcccc@{}}
\toprule
\textbf{Task} & \textbf{Material} & \textbf{Distance (m)} & \textbf{Gain (\%)} & \textbf{Volume (\%)}  \\ \midrule
\textbf{Filtering} &  &  &  &  \\ \cmidrule(lr){1-1}
T1L1 & - & - & 5 & 75 \\
T1L2 & 1 layer foam & - & 5 & 75 \\
T1L3 & T1L2 + 1 layer foam & - & 20 & 60 \\
T1L4 & T1L3 + paper towels + 1 layer foam  & - & 20 & 60 \\
T1L5 & T1L4 + cardboard   & - & 30 & 60 \\
T1L6 & T1L5 + matches   & - & 40 & 60 \\
T1L7 & T1L6  & - & 50 & 50 \\
\textbf{Convolution} &  &  &  &  \\ \cmidrule(lr){1-1}
T2L1 & -  & 1 & 50 & 80 \\
T2L2 & -  & 5 & 55 & 80 \\
T3L3 & -  & 10 & 60 & 80 \\
\textbf{Both} &  &  &  &  \\ \cmidrule(lr){1-1}
T3L1 & T1L2  & 5 & 55 & 80 \\
T3L2 & T1L4 & 10 & 60 & 80 \\
\end{tabular}
\caption{Experimental setup for the different tasks and level. In the material column, referencing another task and level means the same material was used here. In particular, the filter+reverb tasks were recorded first using the reverb parameters of T2L2 and T2L3, then the filter parameters of T1L2 and T1L4. The distance column refers to the distance between the speaker and microphone in the reverb experiments. Note that the gain and volume is given in \% of the max setting as opposed to more descriptive metrics. } 
\label{tab:exp}
\end{table}

\begin{table}[!htb]
\centering
\begin{tabular}{@{}lcccc@{}}
\toprule
\textbf{Task} & \textbf{ID} &\textbf{Total Length(s)} & \textbf{Clean Mean CER} & \textbf{Recorded Mean CER} \\
\midrule
\textbf{Filtering} &  &  &  \\ \cmidrule(lr){1-1}
Level 1  & T1L1 & 2876 & 0.00760 & 0.0419 \\
Level 2  & T1L2 & 2960 & 0.00695 & 0.0772 \\
Level 3  & T1L3 & 2922 & 0.00581 & 0.343 \\
Level 4  & T1L4 & 2974 & 0.00737 & 0.730 \\
Level 5  & T1L5 & 3118 & 0.00739 & 0.910 \\
Level 6  & T1L6 & 2945 & 0.00727 & 0.973 \\
Level 7  & T1L7 & 2975 & 0.00742 & 0.972 \\
\textbf{Reverb} &  &  &  \\ \cmidrule(lr){1-1}
Level 1 & T2L1& 3000 & 0.00817 & 0.126 \\
Level 2 & T2L2& 2643 & 0.00899 & 0.474 \\
Level 3 & T2L3 & 2762 & 0.00962 & 0.557 \\
\textbf{Both} &  &  &  \\ \cmidrule(lr){1-1}
Level 1  & T3L1 & 2643 & 0.00899 & 0.918 \\
Level 2 & T3L2 & 2762 & 0.00962 & 1.00 \\
\bottomrule
\end{tabular}
\caption{Statistics of the data. Note that the T2L2 and T2L3 data is reused for the filter+reverb task. In addition, all reverb data has significantly more padding than the filtering data, meaning that the amount of actual speech is roughly half than that of the filtering data. 
 }
\label{tab:data}
\end{table}

\clearpage

\subsection{Impulse Response}
In addition to the data, we attempted to record some data that can be used to recover the impulse response of the room. For this, we used three separate audio files:
\begin{itemize}
    \item A $30$ second swept sine wave, which is a sine wave that increases in frequency exponentially over the time period. Specifically, this was produced using the Librosa package: \newline \texttt{librosa.chirp(fmin=20, fmax=8000, duration=30, sr=16000)}, and then padding the result.
    \item A $10$ second clip of Gaussian noise with padding.
    \item A short "burst" of Gaussian noise with padding.
\end{itemize}

\begin{figure}[!htb]
\centering
\begin{tabular}{cccc}
\hline
\textbf{ID} & \textbf{Data} & \textbf{ID} & \textbf{Data} \\
\hline
Clean & \includegraphics[width=0.42\textwidth]{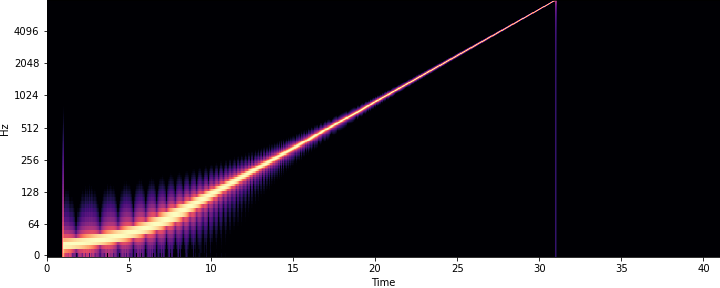} &    T1L1 & \includegraphics[width=0.42\textwidth]{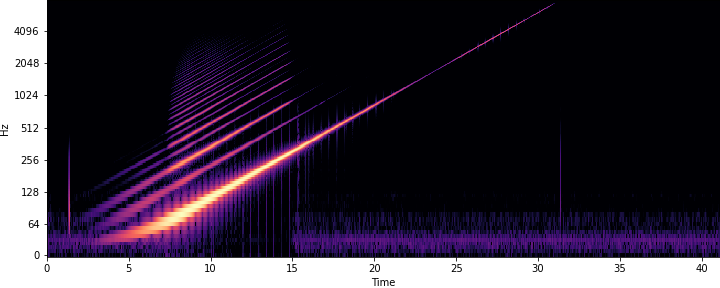} \\
\hline
T1L2 & \includegraphics[width=0.42\textwidth]{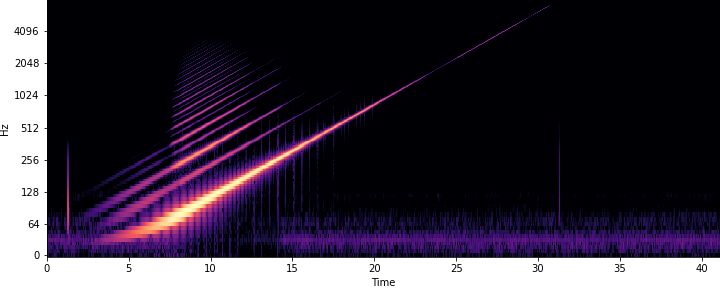} &   T1L3 & \includegraphics[width=0.42\textwidth]{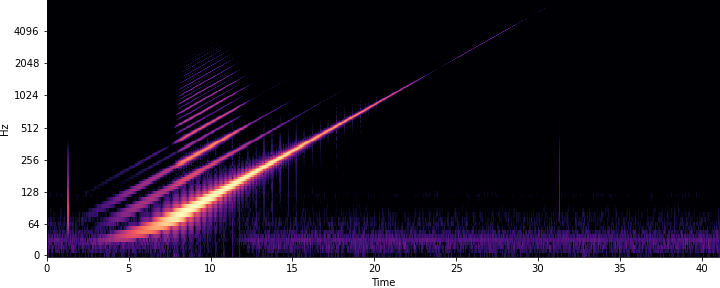} \\
\hline  
T1L4 & \includegraphics[width=0.42\textwidth]{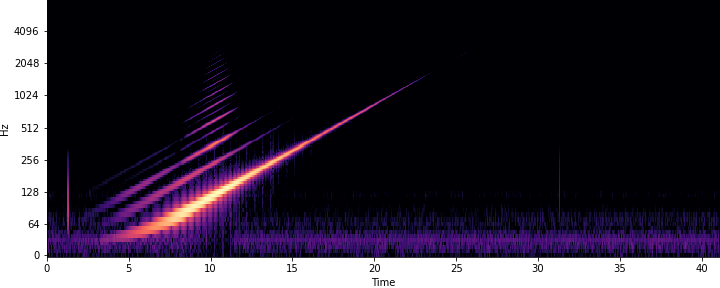} &   T1L5 & \includegraphics[width=0.42\textwidth]{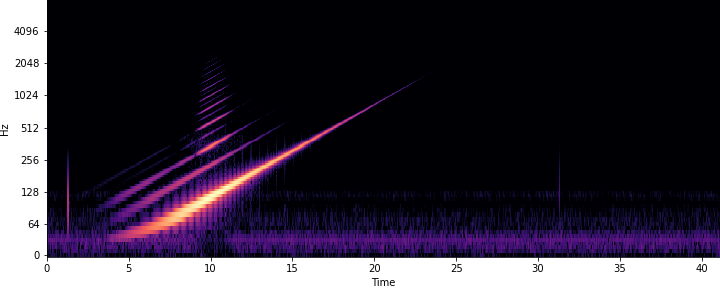} \\
\hline  
T1L6 & \includegraphics[width=0.42\textwidth]{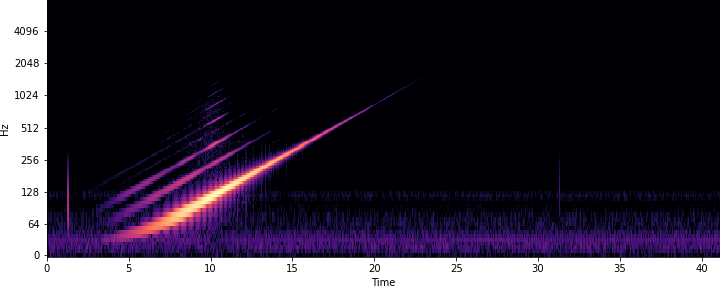} &   T1L7 & \includegraphics[width=0.42\textwidth]{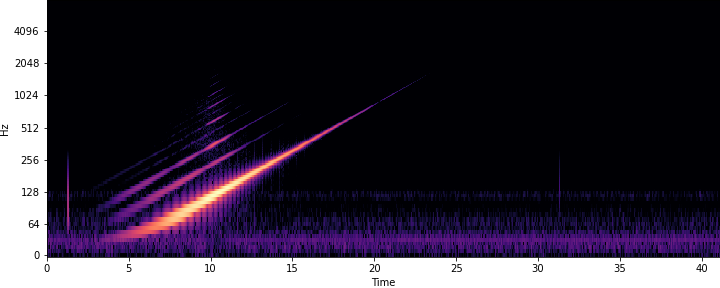} \\
\hline 
T2L1 & \includegraphics[width=0.42\textwidth]{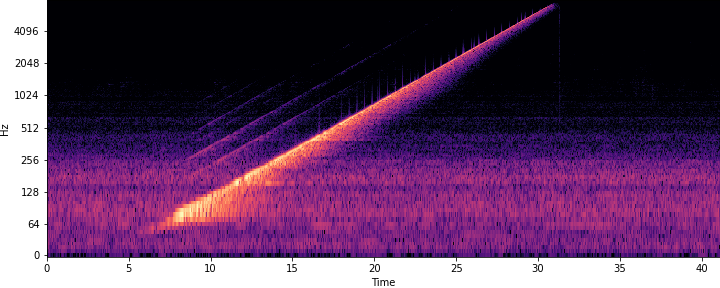} &   T2L2 & \includegraphics[width=0.42\textwidth]{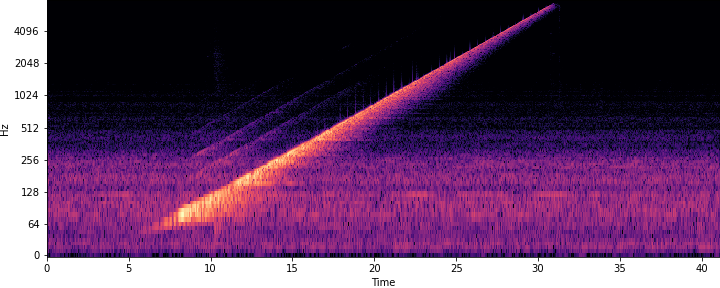} \\
\hline
T2L3 & \includegraphics[width=0.42\textwidth]{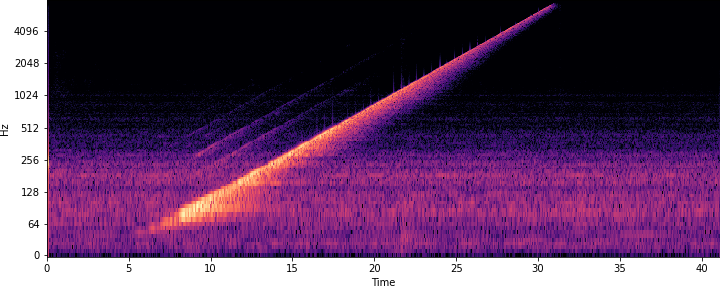} &  \\
\hline
% Add more levels as needed
\end{tabular}
\caption{Clean and recorded IR for different levels. We clearly see that for the filtering experiment, higher frequencies are attenuated as the levels increase. Additionally, there is noticeable non-linear resonance for frequencies between 50 Hz and 250 Hz. The non-linearities are less pronounced in the reverb experiments, but these recordings are also much noisier.}
\label{fig:IR}
\end{figure}

With these three clips, the impulse response of the system can be approximated. For the swept sine wave, the results of measuring the IR of the swept sine wave is shown in Figure \ref{fig:IR}.

From the resulting recordings of the swept sine wave, it is evident that especially the filtering experiment exhibits some non-linear resonances. We identify two main reasons for these non-linearities.

Firstly, the speaker is unable to produce low frequencies accurately, leading instead to an overtone series of these low frequencies. This phenomenon is visible in both experiments. Secondly, in the filter experiment, our setup appears to exhibit harmonic distortion in the frequency range of 50 Hz to 250 Hz. This could be due to room modes or some form of vibration resonance in the setup.

Therefore, a linear time-invariant (LTI) model may not be sufficient to accurately model the corruption of the data. 

\section{Rules and Submission Details}

Firstly, we want to remind participants that the primary goal of this challenge is to enhance speech in difficult settings, using a speech recognition model as a quantitative metric. The rules are crafted with this objective in mind. Due to the data volume and the runtime required for the DeepSpeech model to evaluate submissions, certain constraints are necessary to manage time effectively. We encourage participants to adopt a cooperative spirit and do their best to facilitate the evaluation process. Please view these rules as guidelines designed to ensure fair and efficient assessment, rather than as limitations to be circumvented. 
\subsection{Scoring}
The core evaluation of the participants is as follows:
\begin{enumerate}
\item All participants start with 0 points. To advance to the next level, a group must achieve a Mean CER below 0.3 on the current level. If the Mean CER of the noisy data is already below 0.3, participants must achieve a Mean CER lower than that of the noisy data. Additionally, they must pass a sanity check (see below).
\item Upon completing a level of a specific task, the team gains one point and can proceed to the next level of that task. Note that tasks can be completed independently of each other.
\item The winner is the team with the most points, i.e., the team that completes the most levels. In the event of a tie, the winner will be determined by the average Mean CER across all completed levels.
\end{enumerate}

\textbf{The sanity check} will consist of the judges listening to a few predetermined test files containing speech passed through the algorithm, that may or may not be obtained through a different synthesis process than the training data. \textbf{The groups will only pass the sanity check if it is clear that the recovered audio comes from the same speaker as the clean and noisy data.}

\subsection{Additional Rules}
\begin{itemize}
\item Submissions must adhere to the guidelines outlined in Section \ref{sec:submission}.
\item Each group is allowed a maximum of three submissions.
\item Models must handle 16 kHz audio data of arbitrary length.
\item Participants should primarily use the provided dataset and avoid augmenting it with external data. If additional data is used, it must be explicitly stated, along with results demonstrating its impact on performance. Generating new data from the provided dataset (e.g., creating noisy data from clean data) is permitted. \textbf{Using the OpenAI tts-1 model is not allowed}, as this is a commercial text-to-speech software, and would skew results in favour to those who purchase a subscription.
\item Although participants receive matching text for evaluation, \textbf{they are prohibited from using speech recognition models during training.} This includes optimizing or backpropagating through models like DeepSpeech. Parameter tuning based on the test script output is allowed. Participants are encouraged to explore and report on the effect of backpropagating through speech recognition models outside of the official submission.
\item Participants are encouraged to create lightweight models. \textbf{The Real-Time Factor (RTF), defined as processing time divided by audio length, must average no more than $3$}. All groups' RTFs will be reported. Models achieving an RTF below 1 are particularly encouraged. Evaluation will be conducted on a modern workstation with a GPU. 
\end{itemize}

\subsection{Structure of Provided Files}
The data can be downloaded from Zenodo: \url{https://zenodo.org/records/11380835}.
The files are structured so that each folder contains a folder with clean and folder with recorded data with matching names for the individual files, along with a textfile of the recorded text. \textbf{Note that for \texttt{Task\_3\_Level\_1} and \texttt{Task\_3\_Level\_2}, the clean data is found in \texttt{Task\_2\_Level\_2} and \texttt{Task\_2\_Level\_3} respectively, as these tasks and levels use the same clean data.}

The folder with data also contains a script called \texttt{evaluate.py}, which can be used to test data with the DeepSpeech model. In order to run this script, the participants first need to install and download the Mozilla Deepspeech model. A guide on how to do this is found in its documentation: \url{https://deepspeech.readthedocs.io/en/r0.9/}. We recommend using the v0.9.3 model. Note that Python version 3.5 to 3.9 is required. Additional Python libraries needed include librosa, jiwer, numpy, and pandas. These can be installed using the following pip command.

When it is installed and downloaded, the evaluate script can be called with the following arguments:
\begin{mdframed}
\begin{lstlisting}
python evaluate.py --audio_dir /path/to/audio \
                      --text_file /path/to/text.txt \
                      --output_csv /path/to/output.csv \
                      --model_path /path/to/model.pbmm \
                      --scorer_path /path/to/scorer.scorer \
                      --verbose 1
\end{lstlisting}
\end{mdframed}

The script will then return a .csv files containing the transcription of each audio file. \textbf{Note that the filenames in the audio directory need to match the filenames in the text file.}

As an example, if you download \texttt{Task\_1\_Level\_1.zip}, and unzip it in the same folder as where the models are placed, you can run the recorded data by calling:

\begin{mdframed}
\begin{lstlisting}
python evaluate.py --audio_dir Recorded \
                      --text_file Task_1_Level_1_text_samples.txt \
                      --output_csv output.csv \
                      --model_path deepspeech-0.9.3-models.pbmm \
                      --scorer_path deepspeech-0.9.3-models.scorer \
                      --verbose 1
\end{lstlisting}
\end{mdframed}
The results will then be saved in \texttt{output.csv}.

\subsection{Submission Details}  % THIS SUBSECTION IS OK TO HAVE IN THIS DOCUMENT BUT NOT IN THE FINAL DATA-PAPER
\label{sec:submission}

First, some important dates:
\begin{itemize}
    \item Data Challenge Launch: 10. June 2024.
    \item Sign-up deadline: 1. September 2024 (if you missed this deadline and wish to participate in the challenge, please send us an email). 
    \item Submission deadline: 6. October 2024. We realize this deadline is a bit optimistic, but we humbly ask participants to try to make this deadline.
    \item Results are published: 4. November.
    \item Inverse days: 10.-13. December in Oulu, Finland.
\end{itemize}

The algorithms must be shared with us as a private GitHub repository at latest on the deadline. The codes should be in Matlab or Python3.

After the deadline there is a brief period during which we can troubleshoot the codes together with the participants. This is to ensure that we are able to run the codes.

Participants can update the contents of the shared repository as many times as needed before the deadline. We will consider only the latest release of your repository on Github.

Your repository must contain a README.md file with at least the following sections:

\begin{itemize}
\item Authors, institution, location.
\item Brief description of your algorithm and a mention of the competition.
\item Installation instructions, including any requirements.
\item Usage instructions.
\item An illustration of some example results produced by their model, either audio files or spectrograms.
\end{itemize}

The repository must contain a main routine that we can run to apply your algorithm automatically to every audio file in a given directory, and store the result with the same name in a different folder. This is the file we will run to evaluate your code. Give it an easy to identify name like \texttt{main.m} or \texttt{main.py}.

Your main routine must require three input arguments:
\begin{itemize}
    \item (string) Folder where the input audio files are located.
    \item (string) Folder where the output output audio files will be stored.
    \item (string) task ID on the form \texttt{TXLY}, where \texttt{X} is the task and \texttt{Y} is the level.
\end{itemize}
Below are the expected formats of the main routines in python and Matlab:

Matlab: The main function must be a callable function:
\begin{mdframed}
\begin{lstlisting}
    function main(inputFolder,outputFolder,taskID)
    ...
    
    your code comes here
    ...
\end{lstlisting}
\end{mdframed}
Example calling the function:
\begin{mdframed}
\begin{lstlisting}
>> main('path/to/input/files', 'path/to/output/files', T1L3)
\end{lstlisting}
\end{mdframed}

Python: The main function must be a callable function from the command line. To achieve this you can use sys.argv or argparse module.

Example calling the function:
\begin{mdframed}
\begin{lstlisting}
$ python3 main.py path/to/input/files path/to/output/files T1L3
\end{lstlisting}
\end{mdframed}

The main routine must produce deconvolved audio files in the output folder with the same name for each audio file in the input folder, saved in .wav format. There is no requirement that the audio files are of the exact same length, but they should not be much longer, and they should be 16-bit 16kHz audio files. 

The teams are allowed to use freely available python modules or Matlab toolboxes. Toolboxes, libraries and modules with paid licenses can also be used if the organizing committee also have the license. For example, the most usual Matlab toolboxes for audio processing and deconvolutioncan be used (audio toolbox, wavelet toolbox, PDE toolbox, deep learning toolbox, optimization toolbox). For Python, we recommend using the Librosa package \cite{librosa} and/or PyTorch/Torchaudio \cite{pytorch_article,yang2021torchaudio,hwang2023torchaudio}. The teams can contact us to check if other toolboxes and packages are available.

Finally, the competitors must make their GitHub repositories public at latest on 27. October 2024. In the spirit of open science, only a public code can win the data challenge.

\section*{Acknowledgements}
We would like to extend our heartfelt thanks to the Centre of Excellence of Inverse Modelling and Imaging, the FAME flagship and the Finnish Inverse Problems Society for providing us with the opportunity to create this challenge. 

We are also deeply grateful to Professor Ville Pulkki and the Aalto University Acoustics Lab for their discussions, inspiration, and feedback under the development of the data challenge.

Special thanks go to Leevi Leino, a master's student from Helsinki University, whose contributions to building the experimental setup were important.

Lastly, our appreciation extends to the Inverse Problems group at the University of Helsinki for their insightful discussions and feedback, which greatly enhanced the data collection process.

\bibliographystyle{unsrt} % We choose the "unsrt" reference style, refences in order of appearance 
\bibliography{ref} % Entries are in the refs.bib file

\begin{thebibliography}{10}

\bibitem{mueller2012linear}
Jennifer~L Mueller and Samuli Siltanen.
\newblock {\em Linear and nonlinear inverse problems with practical applications}.
\newblock SIAM, 2012.

\bibitem{speechEnhancement}
Douglas O'Shaughnessy.
\newblock Speech enhancement—a review of modern methods.
\newblock {\em IEEE Transactions on Human-Machine Systems}, 54(1):110--120, 2024.

\bibitem{HDC2021}
Markus Juvonen, Samuli Siltanen, and Fernando~Silva de~Moura.
\newblock Helsinki deblur challenge 2021: Description of photographic data.
\newblock {\em Inverse Problems and Imaging}, 17(5):1008--1023, 2023.

\bibitem{PESQ}
A.W. Rix, J.G. Beerends, M.P. Hollier, and A.P. Hekstra.
\newblock Perceptual evaluation of speech quality (pesq)-a new method for speech quality assessment of telephone networks and codecs.
\newblock In {\em 2001 IEEE International Conference on Acoustics, Speech, and Signal Processing. Proceedings (Cat. No.01CH37221)}, volume~2, pages 749--752 vol.2, 2001.

\bibitem{STOI}
Cees~H. Taal, Richard~C. Hendriks, Richard Heusdens, and Jesper Jensen.
\newblock A short-time objective intelligibility measure for time-frequency weighted noisy speech.
\newblock In {\em 2010 IEEE International Conference on Acoustics, Speech and Signal Processing}, pages 4214--4217, 2010.

\bibitem{DeepSpeech_article}
Awni Hannun, Carl Case, Jared Casper, Bryan Catanzaro, Greg Diamos, Erich Elsen, Ryan Prenger, Sanjeev Satheesh, Shubho Sengupta, Adam Coates, and Andrew Ng.
\newblock Deepspeech: Scaling up end-to-end speech recognition.
\newblock 12 2014.

\bibitem{LibriSpeech}
Vassil Panayotov, Guoguo Chen, Daniel Povey, and Sanjeev Khudanpur.
\newblock Librispeech: An asr corpus based on public domain audio books.
\newblock In {\em 2015 IEEE International Conference on Acoustics, Speech and Signal Processing (ICASSP)}, pages 5206--5210, 2015.

\bibitem{WSJ0-2mix}
John~R. Hershey, Zhuo Chen, Jonathan Le~Roux, and Shinji Watanabe.
\newblock Deep clustering: Discriminative embeddings for segmentation and separation.
\newblock In {\em 2016 IEEE International Conference on Acoustics, Speech and Signal Processing (ICASSP)}, pages 31--35, 2016.

\bibitem{OpenAI_tts}
OpenAI.
\newblock Text to speech.
\newblock \url{https://platform.openai.com/docs/guides/text-to-speech}.
\newblock Accessed: 2024-04-04.

\bibitem{Conrad_gutenberg}
Joseph Conrad.
\newblock {\em The Arrow of Gold: A Story Between Two Notes}.
\newblock \url{https://www.gutenberg.org/ebooks/1083}, 1997.
\newblock Accessed: 2024-03-19.

\bibitem{Austen_gutenberg}
Jane Austen.
\newblock {\em Emma}.
\newblock \url{https://www.gutenberg.org/ebooks/158}, 1994.
\newblock Accessed: 2024-02-16.

\bibitem{Norris_gutenberg}
Kathleen~Thompson Norris.
\newblock {\em Harriet and the Piper}.
\newblock \url{https://www.gutenberg.org/ebooks/5006}, 2004.
\newblock Accessed: 2024-03-19.

\bibitem{James_gutenberg}
Henry James.
\newblock {\em The Turn of the Screw}.
\newblock \url{https://www.gutenberg.org/ebooks/209}, 1995.
\newblock Accessed: 2024-02-29.

\bibitem{Kipling_gutenberg}
Rudyard Kipling.
\newblock {\em The Jungle Book}.
\newblock \url{https://www.gutenberg.org/ebooks/236}, 2006.
\newblock Accessed: 2024-02-29.

\bibitem{Chambers_gutenberg}
Robert~W. Chambers.
\newblock {\em The King in Yellow}.
\newblock \url{https://www.gutenberg.org/ebooks/8492}, 2005.
\newblock Accessed: 2024-03-19.

\bibitem{Collodi_gutenberg}
Carlo Collodi.
\newblock {\em The Adventures of Pinocchio}.
\newblock \url{https://www.gutenberg.org/ebooks/500}, 2006.
\newblock Accessed: 2024-02-29.

\bibitem{Forster_gutenberg}
Edward~Morgan Forster.
\newblock {\em A Room with a View}.
\newblock \url{https://www.gutenberg.org/ebooks/2641}, 2001.
\newblock Accessed: 2024-03-20.

\bibitem{Barclay_gutenberg}
Florence~L. Barclay.
\newblock {\em The Rosary}.
\newblock \url{https://www.gutenberg.org/ebooks/3659}, 2003.
\newblock Accessed: 2024-03-19.

\bibitem{Wells_gutenberg}
Herbert~George Wells.
\newblock {\em The Time Machine}.
\newblock \url{https://www.gutenberg.org/ebooks/35}, 2004.
\newblock Accessed: 2024-02-29.

\bibitem{Speaker_Genelec}
Genelec.
\newblock 8010a studio monitor.
\newblock \url{https://www.genelec.com/8010a#section-technical-specifications}.
\newblock Accessed: 2024-05-28.

\bibitem{microphone}
RØDE.
\newblock RØdelink lav.
\newblock \url{https://rode.com/en/microphones/lavalier-wearable/rodelink-lav}.
\newblock Accessed: 2024-05-28.

\bibitem{recorder}
Zoom.
\newblock H4n pro.
\newblock \url{https://zoomcorp.com/en/us/handheld-recorders/handheld-recorders/h4n-pro/}.
\newblock Accessed: 2024-05-28.

\bibitem{librosa}
Brian McFee, Matt McVicar, Daniel Faronbi, Iran Roman, Matan Gover, Stefan Balke, Scott Seyfarth, Ayoub Malek, Colin Raffel, Vincent Lostanlen, Benjamin van Niekirk, Dana Lee, Frank Cwitkowitz, Frank Zalkow, Oriol Nieto, Dan Ellis, Jack Mason, Kyungyun Lee, Bea Steers, Emily Halvachs, Carl Thomé, Fabian Robert-Stöter, Rachel Bittner, Ziyao Wei, Adam Weiss, Eric Battenberg, Keunwoo Choi, Ryuichi Yamamoto, CJ~Carr, Alex Metsai, Stefan Sullivan, Pius Friesch, Asmitha Krishnakumar, Shunsuke Hidaka, Steve Kowalik, Fabian Keller, Dan Mazur, Alexandre Chabot-Leclerc, Curtis Hawthorne, Chandrashekhar Ramaprasad, Myungchul Keum, Juanita Gomez, Will Monroe, Viktor~Andreevitch Morozov, Kian Eliasi, nullmightybofo, Paul Biberstein, N.~Dorukhan Sergin, Romain Hennequin, Rimvydas Naktinis, beantowel, Taewoon Kim, Jon~Petter Åsen, Joon Lim, Alex Malins, Darío Hereñú, Stef van~der Struijk, Lorenz Nickel, Jackie Wu, Zhen Wang, Tim Gates, Matt Vollrath, Andy Sarroff, Xiao-Ming, Alastair Porter, Seth Kranzler, Voodoohop,
  Mattia~Di Gangi, Helmi Jinoz, Connor Guerrero, Abduttayyeb Mazhar, toddrme2178, Zvi Baratz, Anton Kostin, Xinlu Zhuang, Cash~TingHin Lo, Pavel Campr, Eric Semeniuc, Monsij Biswal, Shayenne Moura, Paul Brossier, Hojin Lee, and Waldir Pimenta.
\newblock librosa/librosa: 0.10.2.post1.
\newblock \url{https://doi.org/10.5281/zenodo.11192913}, May 2024.

\bibitem{pytorch_article}
Adam Paszke, Sam Gross, Francisco Massa, Adam Lerer, James Bradbury, Gregory Chanan, Trevor Killeen, Zeming Lin, Natalia Gimelshein, Luca Antiga, Alban Desmaison, Andreas Kopf, Edward Yang, Zachary DeVito, Martin Raison, Alykhan Tejani, Sasank Chilamkurthy, Benoit Steiner, Lu~Fang, Junjie Bai, and Soumith Chintala.
\newblock Pytorch: An imperative style, high-performance deep learning library.
\newblock In {\em Advances in Neural Information Processing Systems 32}, pages 8024--8035. Curran Associates, Inc., 2019.

\bibitem{yang2021torchaudio}
Yao-Yuan Yang, Moto Hira, Zhaoheng Ni, Anjali Chourdia, Artyom Astafurov, Caroline Chen, Ching-Feng Yeh, Christian Puhrsch, David Pollack, Dmitriy Genzel, Donny Greenberg, Edward~Z. Yang, Jason Lian, Jay Mahadeokar, Jeff Hwang, Ji~Chen, Peter Goldsborough, Prabhat Roy, Sean Narenthiran, Shinji Watanabe, Soumith Chintala, Vincent Quenneville-Bélair, and Yangyang Shi.
\newblock Torchaudio: Building blocks for audio and speech processing.
\newblock {\em arXiv preprint arXiv:2110.15018}, 2021.

\bibitem{hwang2023torchaudio}
Jeff Hwang, Moto Hira, Caroline Chen, Xiaohui Zhang, Zhaoheng Ni, Guangzhi Sun, Pingchuan Ma, Ruizhe Huang, Vineel Pratap, Yuekai Zhang, Anurag Kumar, Chin-Yun Yu, Chuang Zhu, Chunxi Liu, Jacob Kahn, Mirco Ravanelli, Peng Sun, Shinji Watanabe, Yangyang Shi, Yumeng Tao, Robin Scheibler, Samuele Cornell, Sean Kim, and Stavros Petridis.
\newblock Torchaudio 2.1: Advancing speech recognition, self-supervised learning, and audio processing components for pytorch, 2023.

\end{thebibliography}

% \section{References}
% (Make bibtex)
% \begin{itemize}
%     \item https://www.genelec.com/8010a#section-technical-specifications
%     \item https://rode.com/en/microphones/lavalier-wearable/rodelink-lav
%     \item https://zoomcorp.com/en/us/handheld-recorders/handheld-recorders/h4n-pro/
% \end{itemize}

\section*{Appendix}

\begin{table}[h!]
\centering
\begin{tabular}{@{}lccc@{}}
\toprule
\textbf{Task} & \textbf{Number of samples} & \textbf{Voice distribution} & \textbf{Sentence length distribution} \\
\midrule
\textbf{Filtering} &  &  &  \\ \cmidrule(lr){1-1}
Level 1  & 600 & (100, 104, 93, 102, 102, 99) & (290, 288, 22) \\
Level 2  & 611 & (107, 100, 101, 99, 102, 102) & (299, 287, 25)\\
Level 3  & 611 & (102, 106, 103, 99, 100, 101) & (303, 283, 25) \\
Level 4  & 611 & (102, 104, 97, 105, 100, 103) & (296, 291, 24) \\
Level 5  & 611 & (108, 104, 101, 99, 100, 99) & (295, 291, 25) \\
Level 6  & 611 & (100, 102, 103, 102, 102, 102) & (295, 292, 24)  \\
Level 7  & 611 & (105, 107, 98, 101, 99, 101) & (292, 295, 24) \\
\textbf{Reverb} &  &  &  \\ \cmidrule(lr){1-1}
Level 1  & 323 & (60, 50, 52, 56, 49, 56) & (166, 145, 12) \\
Level 2  & 280 & (43, 51, 43, 46, 48, 49) & (132, 136, 12) \\
Level 3 & 296 & (45, 53, 49, 52, 47, 50) & (142, 144, 10) \\
\bottomrule
\end{tabular}
\caption{We used OpenAI text-to-speech model to generate clean data for the challenge. The model contains six build in voices. The used voices in each level are presented in the column "Voice distribution" in the following order ("Alloy", "Echo", "Fable", "Onyx", "Nova", "Shimmer"). The speech samples were generated using text samples of different lengths. The used text samples' length by number of words are presented in the column "Sentence length distribution", where the first value is the number of short sentences (6-10 words), the second value is the number of medium length sentences (11-20 words) and the last value is the number of long sentences ($>20$ words).}
\label{tab:data_info_clean}
\end{table}

\end{document}